\documentclass{article}
\usepackage{spconf,amsmath,graphicx,hyperref}
\usepackage{booktabs}

\title{LiveAnon: Improving Privacy and Utility in Streaming Speaker Anonymization with Codec Language Models}
\name{Yejin Kim$^*$, Kyungsu Kim$^*$, Seungu Han$^*$, Kyogu Lee$^{* \dag \ddagger}$}
\address{$^*$ Music and Audio Research Group (MARG)\\
  $^\dag$ Department of Intelligence and Information, Seoul National University\\
  $^\ddagger$ Interdisciplinary Program in Artificial Intelligence, Seoul National University}
\begin{document}
\maketitle
\begin{abstract}
Speaker anonymization (SA) aims to conceal speaker identity while preserving linguistic content and emotion, but existing streaming systems often lose emotion information through discrete content bottlenecks or require emotion-specific modeling.
We present \textit{LiveAnon}, a voice-conversion-based streaming SA system that directly conditions an autoregressive codec language model on hidden features from a pretrained streaming ASR encoder, preserving emotion without emotion-specific supervision.
To reduce codec-generation latency, we introduce capped RVQ decoding, which preserves dependencies among coarse codebooks while predicting the remaining codebooks in parallel.
Under VPC24, LiveAnon improves intelligibility and emotion preservation over fixed-target \textit{Stream-Voice-Anon (SVA)}, a strong prior streaming SA system, matches emotion-prompted SVA in emotion preservation, and achieves the strongest semi-informed privacy among streaming systems.
Under VPC26, it ranks second across all metrics among the evaluated anonymization systems, demonstrating a strong privacy–utility balance.
\end{abstract}
\begin{keywords}
streaming speaker anonymization, privacy preservation, voice conversion, neural audio codec
\end{keywords}
\section{Introduction}
\label{sec:intro}

Speaker anonymization (SA) aims to conceal speaker identity while retaining linguistic and paralinguistic information such as emotion \cite{tomashenko2022voiceprivacy}.
Since these attributes are jointly encoded in speech, suppressing speaker identity often comes at the cost of preserving other speech information, resulting in a privacy-utility trade-off, where utility encompasses speech intelligibility and emotion preservation~\cite{zhang2023voicepm,cai2024privacy}.
Beyond offline processing, streaming anonymization is increasingly important for privacy-sensitive interactive applications, including call centers, voice assistants, medical conversations, and legal communications \cite{meyer2025use}.
Together, these considerations highlight two key challenges for streaming SA: balancing latency with quality and privacy with utility.

Existing SA systems generally follow two main approaches: cascaded ASR-to-TTS \cite{meyer2022speaker, meyer2023anonymizing} and voice conversion (VC)-based anonymization \cite{panariello2024speaker, yao2024npu, huang2024diffvc+, kuzmin2026stream}, with neural VC underlying most state-of-the-art systems \cite{miao2026voiceprivacy}.
VC-based anonymization systems commonly rely on intermediate content representations to separate source content from speaker identity.
Discrete or quantized bottlenecks can suppress speaker information, but may also discard information important for emotion preservation~\cite{gaznepoglu2025disentanglement}.
These challenges are further compounded in streaming SA, where anonymization must operate with limited future context and bounded latency.

Preserving emotion is important for maintaining the utility of anonymized speech \cite{tomashenko2024voiceprivacy}.
Various approaches have therefore incorporated explicit prosodic cues \cite{meyer2023prosody, franzreb2025private, he2025emotion} or emotion representations distilled from pretrained models \cite{yao2025easy, kuzmin2026streamvoiceanon+} to improve emotion preservation.
However, such approaches require additional components for prosody or emotion modeling in the anonymization pipeline.
This motivates investigating whether richer paralinguistic information can instead be retained directly in the source content representation.

To this end, we propose \textit{LiveAnon}, a VC-based streaming SA framework using an autoregressive codec language model.
LiveAnon directly conditions acoustic generation on hidden representations from a pretrained streaming Zipformer~\cite{yao2024zipformer}.
Our analysis shows that these representations retain substantial emotion-related information while exhibiting less speaker discriminability.
Target identity is specified through prefix codec prompt, in which a reference codec sequence is directly continued within the same autoregressive acoustic stream without requiring an explicit speaker encoder.
To reduce codec generation latency, we modify MusicGen's delay pattern \cite{copet2023simple} with a capped delay scheme that limits the delay to three codec frames and parallelizes the remaining RVQ codebooks.
LiveAnon improves across all evaluation metrics under the VPC24 \cite{tomashenko2024voiceprivacy} fixed-target setting and remains competitive with the offline VPC26 \cite{miao2026voiceprivacy} baselines while operating in a streaming setting.

The demo is available on: \url{https://jin1025.github.io/LiveAnon/}.

\begin{figure*}[t]
  \centering
  \includegraphics[width=0.9\linewidth]{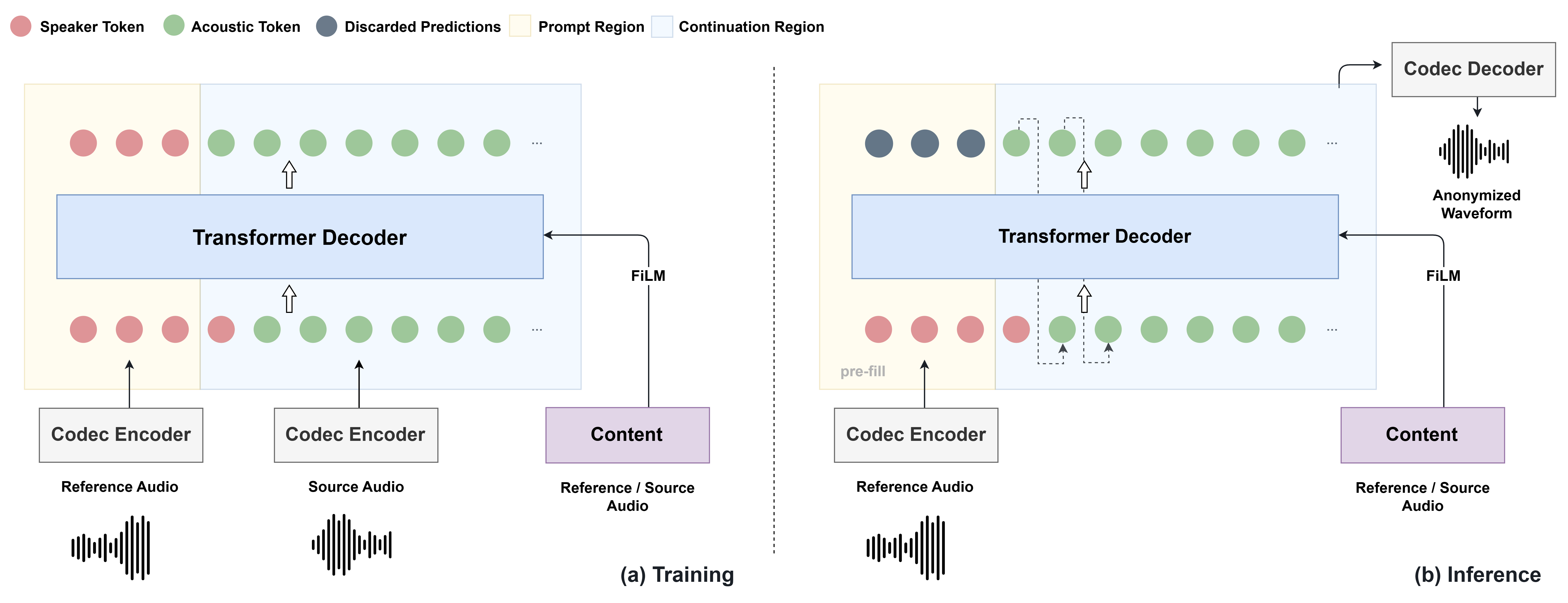}
  \caption{Overview of LiveAnon during (a) training and (b) inference. A 3-s reference utterance provides codec tokens for acoustic prompting, and features from a frozen streaming ASR encoder provide content conditioning. Training uses teacher-forced codec sequences, whereas inference pre-fills the reference prompt into the Transformer KV cache and continues autoregressive generation with the capped RVQ delay pattern.}
  \label{fig:architecture}
\end{figure*}

\section{Method}
\label{sec:method}

\subsection{System Overview}
\label{ssec:overview}

Fig.~\ref{fig:architecture} illustrates the overall architecture of LiveAnon.
The system consists of a pretrained streaming ASR encoder, a neural audio codec, and a causal Transformer decoder.
Given source speech and a reference utterance from the target speaker, the ASR encoder extracts content features from the source, while the codec converts the reference into discrete acoustic tokens that serve as the speaker prompt.
The decoder autoregressively continues this prompt conditioned on the source content, and the generated tokens are reconstructed into a waveform by the codec decoder.

\subsection{Pretrained ASR Content Representation}
\label{ssec:content}
We use hidden representations from a frozen pretrained streaming Zipformer ASR encoder\footnote{We use the streaming Zipformer checkpoint available at \href{https://huggingface.co/Zengwei/icefall-asr-librispeech-streaming-zipformer-2023-05-17/tree/main/exp}{Hugging Face}.} as the source content representation.
We extract the output of the final encoder stage before the final downsampling, preserving a 50-Hz frame rate that matches the codec tokens.
The features are linearly projected to the decoder conditioning dimension.
A fixed temporal shift aligns the ASR and codec representations, and its resulting lookahead is included in the algorithmic latency reported in Sec.~\ref{ssec:latency}.
The aligned content sequence conditions every Transformer layer through feature-wise linear modulation (FiLM) \cite{perez2018film}, where the content features predict per-layer affine modulation parameters.

\subsection{Codec Prompt Continuation and Acoustic Generation}
\label{ssec:codec_prompt}

We use JHCodec \cite{lee2026reconstruct}, a streaming neural audio codec that represents speech with discrete acoustic tokens.
Following VALL-E \cite{wang2023neural}, we adapt acoustic prompting to voice-to-voice anonymization by conditioning codec continuation on source content features.
Rather than using the reference as a separate conditioning prefix, the reference and generated codec tokens form a single autoregressive acoustic stream with shared codebook embeddings.
Each region is conditioned on its aligned content features.
With $\mathbf{c}_n$ denoting the aligned feature, we optimize codec-token prediction over the full sequence with cross-entropy:

\begin{equation}
\mathcal{L}_{\mathrm{CE}}
=
\sum_{n=1}^{N}\sum_{d=1}^{D}
-\log p_{\theta}
\left(
a_{n,d}
\mid
\mathcal{H}_{n,d},
\mathbf{c}_{n}
\right),
\end{equation}
where $a_{n,d}$ is the codec token at frame $n$ and RVQ depth $d$, $N$ is the total number of codec frames, and $D$ is the number of RVQ codebooks.
$\mathcal{H}_{n,d}$ denotes the codec tokens available when predicting $a_{n,d}$ under the RVQ delay pattern.

During training, both regions are teacher-forced, allowing the model to learn acoustic continuation from the reference voice while following the source content.

At inference, the reference codec tokens and their content features are pre-filled into the Transformer KV cache, after which autoregressive generation continues from the end of the prompt using source content.
The prompt is therefore processed only once and does not contribute to the per-frame algorithmic latency. Generated RVQ tokens are decoded into waveform by the JHCodec decoder.

\subsection{Capped RVQ Decoding}
\label{ssec:capped_delay}

To capture dependencies across RVQ codebooks, MusicGen uses a delay pattern in which codebook $k$ is shifted by $k$ frames.
However, the resulting maximum delay increases with codebook depth.
Building on its partial delay variant, we introduce a capped delay that preserves staggered dependencies up to a maximum offset $\delta_{\max}$:
\begin{equation}
    d_k = \min(k,\delta_{\max}), \qquad k=0,\ldots,D-1
\end{equation}

Thus, $\delta_{\max}=0$ gives fully parallel codebook prediction, whereas $\delta_{\max}=D-1$ recovers the original full-delay pattern.
Intermediate values provide a controllable trade-off between inter-codebook dependency and generation latency.
For codec frame duration $\Delta$, the resulting codebook latency is $\delta_{\max}\Delta$.

\section{Experimental Setup}
\label{sec:experiments}

\subsection{Dataset}
\label{ssec:dataset}
LiveAnon is trained on the train-clean-100 and train-clean-360 splits of LibriTTS ($\sim$460~h) \cite{zen2019libritts}, resampled to 16~kHz. We retain utterances of at least 6~s and split each into two contiguous 3~s segments, using the first as the reference prompt and the second as the target continuation.

\subsection{Evaluation Protocol}
We evaluate LiveAnon following the VoicePrivacy Challenge 2024 (VPC24) \cite{tomashenko2024voiceprivacy} and 2026 (VPC26) \cite{miao2026voiceprivacy} Track~1 protocols.
We report word error rate (WER$\downarrow$) for intelligibility, unweighted average recall (UAR$\uparrow$) for emotion preservation, and equal error rate (EER$\uparrow$) for speaker privacy.
Following the VoicePrivacy evaluation protocols, WER and EER are computed on LibriSpeech \cite{panayotov2015librispeech}, while emotion UAR is evaluated on IEMOCAP \cite{busso2008iemocap}.
For both protocols, we report EER with lazy-informed (EER-L) attackers trained on original speech and semi-informed (EER-S) attackers trained on anonymized speech.
Because VPC26 uses a stronger attacker pretrained on large-scale voice-converted speech, its EER values are not directly comparable with those from VPC24 and are reported separately.
For anonymization, we adopt the fixed-target strategy of SVA's vctk-1fix setting \cite{kuzmin2026stream}, using a single utterance from a fixed VCTK \cite{yamagishi2019cstr} speaker as the target prompt for all evaluation utterances. We crop the reference utterance to 3\,s for codec prompting.

\subsection{Model Configuration}
\label{ssec:config}
JHCodec operates at 16\,kHz and 50\,fps, producing 8 RVQ codebooks with a codebook size of 1024.
Content features are extracted from a frozen streaming Zipformer ASR encoder trained on LibriSpeech.
We extract the 512-dimensional representation before the final temporal downsampling, retaining a 50-fps frame rate, and project it to 256 dimensions.
The Zipformer operates with a 16-frame (320-ms) streaming chunk.
To align the content and codec streams, the content features are shifted forward by 6 frames.

The autoregressive codec decoder consists of 12 Transformer layers with a hidden dimension of 768, 8 attention heads, a feed-forward dimension of 3072, and a dropout rate of 0.1, using causal self-attention with ALiBi positional biases.
We use a 3~s codec prompt and set the capped RVQ delay to $\delta_{\max}=3$.
The model is trained for 400k optimization steps on a single RTX 3080 Ti GPU using AdamW with a peak learning rate of $10^{-4}$ and a cosine decay schedule.

\section{Results and Discussion}
\label{sec:results}

\subsection{Content Representation Analysis}
\label{ssec:content_analysis}

To examine the information retained by the content representations, we perform linear probing for speaker identity and emotion.
For both LiveAnon and SVA, frame-level features are summarized using mean and standard-deviation pooling and classified with logistic regression.
Speaker information is evaluated on the held-out LibriSpeech dev-clean split using 40 speakers.
For each speaker, non-overlapping utterances are split into 80\% for training and 20\% for testing.
Emotion information is evaluated on IEMOCAP using the speaker-independent five-fold split of the VoicePrivacy protocol.

\begin{table}[t]
\centering
\caption{Linear probing of content representations.
Lower speaker accuracy and higher emotion UAR are preferred.}
\label{tab:content_probe}
\smallskip
{\small
\begin{tabular}{@{}lcc@{}}
\toprule
\textbf{System} &
\textbf{Speaker Acc.$\downarrow$} &
\textbf{Emotion UAR$\uparrow$} \\
\midrule
SVA & 32.50 & 39.95 \\
LiveAnon & \textbf{28.75} & \textbf{55.17} \\
\bottomrule
\end{tabular}}
\end{table}

As shown in Table~\ref{tab:content_probe}, LiveAnon's ASR representation achieves higher emotion UAR than SVA's content representation (55.17 vs.\ 39.95), while yielding lower speaker classification accuracy (28.75 vs.\ 32.50). These results suggest that LiveAnon's ASR representation retains richer emotion-related information while being less speaker-discriminative than SVA's content representation.

\subsection{Speaker Anonymization Performance}
\label{ssec:anon_results}

\begin{table}[t]
\centering
\caption{Comparison with streaming SA systems under the VPC24 protocol.}
\label{tab:vpc24}
\smallskip
{\small
\begin{tabular}{@{}lcccc@{}}
\toprule
\textbf{System} &
\textbf{WER$\downarrow$} &
\textbf{UAR$\uparrow$} &
\textbf{EER-L$\uparrow$} &
\textbf{EER-S$\uparrow$} \\
\midrule
Original & 1.83 & 70.07 & 5.16 & -- \\
\midrule
DarkStream \cite{quamer2025darkstream} & 8.75 & 34.73 & 47.26 & \underline{21.83} \\
TVTSyn \cite{quamer2026tvtsyn} & 5.35 & 37.32 & 47.55 & 14.57 \\
SVA \cite{kuzmin2026stream} \textit{cremad-emo-4rnd}
    & 6.59 & \underline{44.59} & 46.53 & 18.63 \\
SVA \cite{kuzmin2026stream} \textit{vctk-1fix}
    & \underline{4.54} & 39.71 & 47.19 & 15.92 \\
SVA+ \cite{kuzmin2026streamvoiceanon+} \textit{frame-distill}
    & 5.77 & \textbf{49.22} & \textbf{48.98} & 18.30 \\
\midrule
\textbf{LiveAnon} \textit{vctk-1fix}
    & \textbf{4.31} & 44.09 & \underline{48.75} & \textbf{32.28} \\
\bottomrule
\end{tabular}}
\end{table}

Table~\ref{tab:vpc24} compares LiveAnon with prior streaming SA systems under the VPC24 protocol.
Under the matched \textit{vctk-1fix} setting, LiveAnon outperforms SVA across all metrics, with a particularly large gain in EER-S (32.28 vs.\ 15.92), while also achieving lower WER (4.31 vs.\ 4.54).
LiveAnon also achieves an emotion UAR of 44.09, comparable to SVA's emotion-prompted \textit{cremad-emo-4rnd} setting (44.59), despite using only a single fixed VCTK prompt.
SVA+ attains the highest UAR through frame-level emotion distillation, whereas LiveAnon requires no emotion-specific training supervision and achieves lower WER and substantially higher EER-S.

\begin{table}[t]
\centering
\caption{Comparison with VPC26 challenge baselines under the VPC26 evaluation protocol.}
\label{tab:vpc26}
\smallskip
{\small
\begin{tabular}{@{}lcccc@{}}
\toprule
\textbf{System} &
\textbf{WER$\downarrow$} &
\textbf{UAR$\uparrow$} &
\textbf{EER-L$\uparrow$} &
\textbf{EER-S$\uparrow$} \\
\midrule
Original & 1.82 & 70.07 & 4.75 & -- \\
\midrule
B2 & 9.36 & \textbf{54.59} & 6.89 & 3.98 \\
B3 & \textbf{4.21} & 38.09 & 22.55 & 15.28 \\
B4 & 6.03 & 42.38 & 17.19 & 13.21 \\
B5 & 4.68 & 38.46 & \textbf{28.30} & \textbf{18.55}  \\
\midrule
\textbf{LiveAnon} & \underline{4.30} & \underline{44.09} & \underline{26.08} & \underline{17.99} \\
\bottomrule
\end{tabular}}
\end{table}

Table~\ref{tab:vpc26} compares LiveAnon with the offline VPC26 challenge baselines under the VPC26 evaluation protocol.
LiveAnon achieves the second-best result across all four metrics: the second-lowest WER and the second-highest UAR, EER-L, and EER-S among the compared anonymization systems.
These results show that LiveAnon maintains a competitive and well-balanced privacy--utility performance against offline VPC26 baselines, while operating in a streaming setting.

\subsection{Latency and RTF}
\label{ssec:latency}

We compute algorithmic latency as the sum of the content-encoder chunk
duration, content-alignment lookahead, and codebook delay:
\begin{equation}
L_{\mathrm{alg}}
=
L_{\mathrm{chunk}}
+
L_{\mathrm{align}}
+
L_{\mathrm{codec}}
\end{equation}

\begin{table}[t]
\centering
\caption{Algorithmic latency and real-time factor (RTF) of LiveAnon.}
\label{tab:latency}
\smallskip
{\small
\begin{tabular}{@{}ccccc@{}}
\toprule
\textbf{Chunk} &
\textbf{Align.} &
\textbf{Codec} &
\textbf{Total} &
\textbf{RTF$\downarrow$} \\
\midrule
160 ms & 120 ms & 60 ms & 340 ms & 0.34 \\
320 ms & 120 ms & 60 ms & 500 ms & 0.17 \\
\bottomrule
\end{tabular}}
\end{table}

With the default 320~ms Zipformer chunk, LiveAnon has an algorithmic latency of 500~ms and achieves an RTF of 0.17 on an RTX 4090.
The content encoder constitutes the largest latency component.
We additionally evaluate a 160~ms chunk configuration, which reduces the total algorithmic latency to 340 ms.
This suggests that replacing the current Zipformer with a content encoder designed for shorter streaming chunks could further reduce latency.

\subsection{Capped Delay Trade-off}
\label{ssec:tradeoff}

\begin{figure}[t]
    \centering
    \includegraphics[width=\columnwidth]{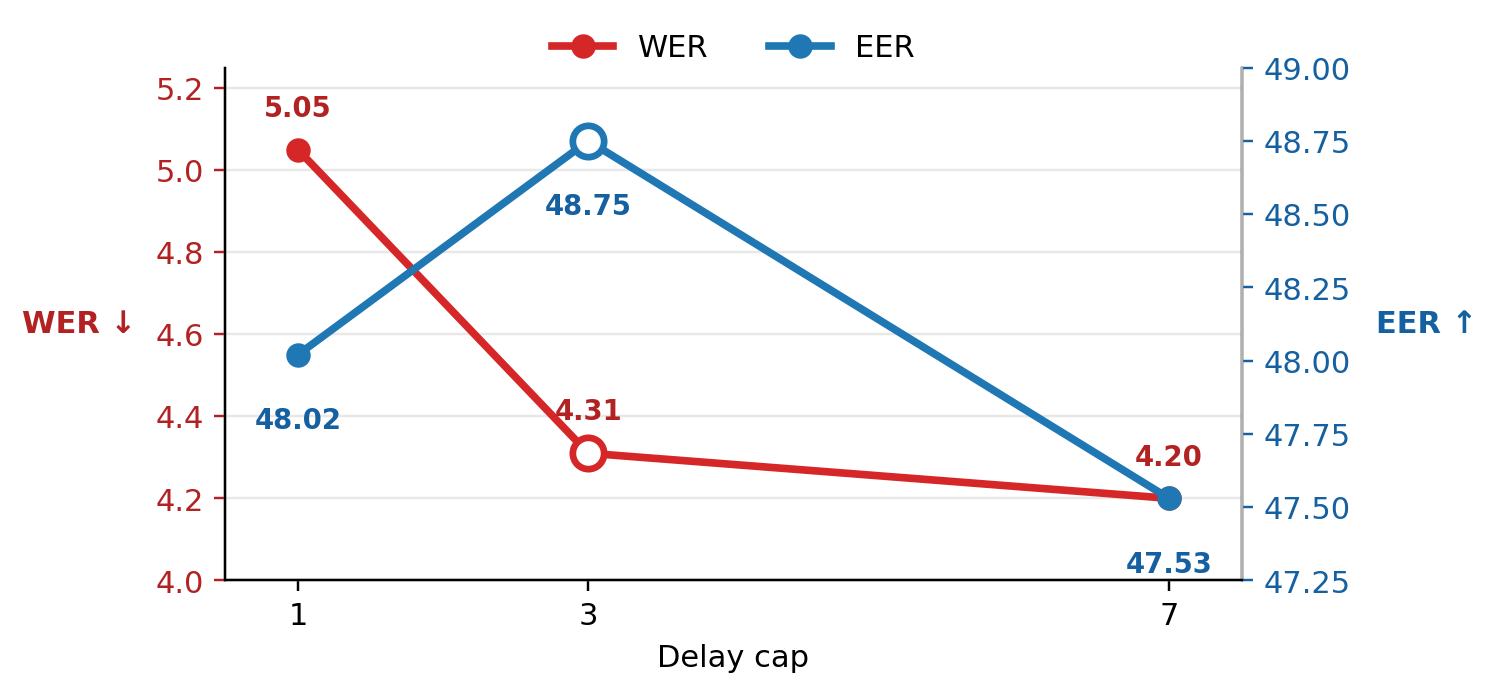}
    \caption{Privacy--utility--latency trade-off under different capped-delay settings. Delay cap 3 provides the best operating point.}
    \label{fig:delay_cap_tradeoff}
\end{figure}

We conduct an ablation study on the delay cap $\delta_{\max}$ to identify its privacy--utility--latency trade-off, using a 320\,ms chunk size.
We evaluate $\delta_{\max}\in\{1, 3, 7\}$, corresponding to codec latencies of 20, 60, and 140\,ms, respectively.
As shown in Figure~\ref{fig:delay_cap_tradeoff}, increasing the delay cap from 1 to 3 improves both privacy and utility.
Further increasing the cap to 7 provides only a marginal utility improvement, while degrading privacy and increasing latency.
Therefore, $\delta_{\max}=3$ (60\,ms codec latency) provides the best balance between latency, privacy, and utility, and is used for all experiments unless otherwise stated.

\section{Conclusion and Future Work}
\label{ssec:anon_conclusion}

We presented LiveAnon, a streaming speaker anonymization system that directly conditions autoregressive codec generation on representations from a pretrained streaming ASR encoder.
LiveAnon preserves emotion-related information without emotion-specific data or supervision, while capped RVQ decoding balances codec latency and generation quality.
On VPC24, LiveAnon leads streaming systems in WER and EER-S and matches emotion-prompted SVA in UAR without emotion-specific modeling, while remaining competitive with offline VPC26 baselines.
While our fixed-target setting follows the VPC protocol, prompt diversity and cross-utterance linkability remain unexplored.
Exploring diverse per-utterance target prompts and lower-latency streaming ASR encoders remains for future work, as the current latency largely depends on the encoder chunk size.

\ninept
\bibliographystyle{IEEEbib}
\bibliography{references}

\end{document}